**Expand or better manage protected areas: a framework for minimising extinction risk when threats are concentrated near edges**

*Brendan G Dillon[1,2], Hugh P Possingham[1,2] and Matthew H Holden[2,3]*

School of the Environment, The University of Queensland, St Lucia, Australia, 4072

Centre for Biodiversity and Conservation Science, The University of Queensland, St Lucia, Australia, 4072

School of Mathematics and Physics, The University of Queensland, St Lucia, Australia, 4072

Abstract

Several international agreements have called for the rapid expansion of protected areas to halt biodiversity declines. However, recent research has shown that expanding protected areas may be less cost-effective than redirecting resources towards threat management in existing reserves. These findings often assume that threats are homogeneously distributed in the landscape. In some cases, threats are more concentrated near the edge of protected areas. As protected areas expand, core habitat in the centre expands more rapidly than its edge, potentially creating a refuge from threats. In this paper, we present a framework linking protected area expansion and threat management to extinction risk, via their impact on population carrying capacity and growth rate within core and edge habitats. We demonstrate the framework using a simple population model where individuals are uniformly distributed in a circular protected area threatened by poachers who penetrate the protected area to a fixed distance. We parameterize the model for Peter's Duiker (*Cephalophus callipygus*) harvested for food in the dense undergrowth of African forests using snares. Expanding protected areas can reduce extinction risk more effectively compared to an equivalent investment in snare removal for larger protected areas that already sustain core unhunted habitat. Our results demonstrate the importance of protected area expansion in buffering susceptible populations from fixed hunting pressure restricted to protected area edges. However, for cases where threats, wildlife, and managers respond to each other strategically in space, the relative importance of expansion versus increased management remains an important open problem.

## Introduction

Despite the impressive expansion of the global protected area estate, species population declines, and extinction risk remain largely unabated (WWF 2024). This is likely because few reserves have the resources to stop all threats to the populations they aim to protect (Pressey et al., 2007). Without adequate investment in threat management, protected areas are little more than "paper parks" (Eyre, 1990, Brandon et al., 1998).

Due to the inadequacies of the current protected area estate, successive Conferences of the Parties (COP) to the Convention on Biological Diversity have increasingly emphasised the importance of protected area management and extinction risk reduction. Target 3 of the Kunming-Montreal Global Biodiversity Framework (GBF) pledges that "…by 2030 at least 30 per cent of terrestrial, inland water, and of coastal and marine areas… *are effectively conserved and managed…"* (CBD, 2022). Such an expansion of the protected area estate promises to reduce the extinction risk of threatened species – *if, and only if,* it is adequately managed. Even as some threatening processes are curtailed by protected area gazettement, numerous others remain active within protected areas and continue to impact populations (Kearney et al. 2020). The choice of whether to deploy limited conservation resources on further expanding protected areas or better managing them is essential to minimising the rate of population extinction.

Recent work addressing the trade-off between protected area expansion and management suggest that the optimal strategy is to invest more resources towards management and restoration of existing protected areas (Kuempel et al., 2018, Adams et al., 2019, Timms and Holden, 2024) rather than expanding them. However, these studies link the expand/manage actions to population size (Kuempel et al., 2018, Timms and Holden, 2024) or species area relationships (Adams et al., 2019) rather than extinction risk, and assume the spatial threats to populations are uniform through protected areas. Numerous threatening processes are most intense at the edges of protected areas rather than uniform in space (Laurance and Yensen 1991). Such edge effects have been seen in patterns of illegal poaching and the imperilment of both mammals (Woodroffe and Ginsberg, 1998) and fish (Ohaun et al., 2021). Even high profile protected areas, including UNESCO World Heritage Sites, face growing human pressures at their perimeter (Wittemyer et al., 2008) and within their boundaries (Allan et al., 2017). One approach to isolating populations from the processes that threaten populations in protected area edges is to establish protected areas that are large enough and round enough to retain a substantial core of habitat unaffected by the threat (UNESCO 1996). Given the recognized importance of such threat-free core habitat, it is essential that prioritization between protected area expansion and threat management incorporates edge effects.

In this paper, we investigate the implications of edge effects for the decision of whether to expand or better manage protected areas. To do so, we use a model of extinction risk which connects a species' intrinsic population growth rate (which is reduced by threats) and carrying capacity (which increases with area) to a populations' mean time to extinction. We then explore the trade-off between increases in habitat area and threat management for reducing extinction risk (Fig. 1). Finally, we apply the model to a case study of Peter's Duiker (*Cephalophus callipygus*), a forest antelope hunted with snares in Central Africa, to evaluate when each strategy is most effective.

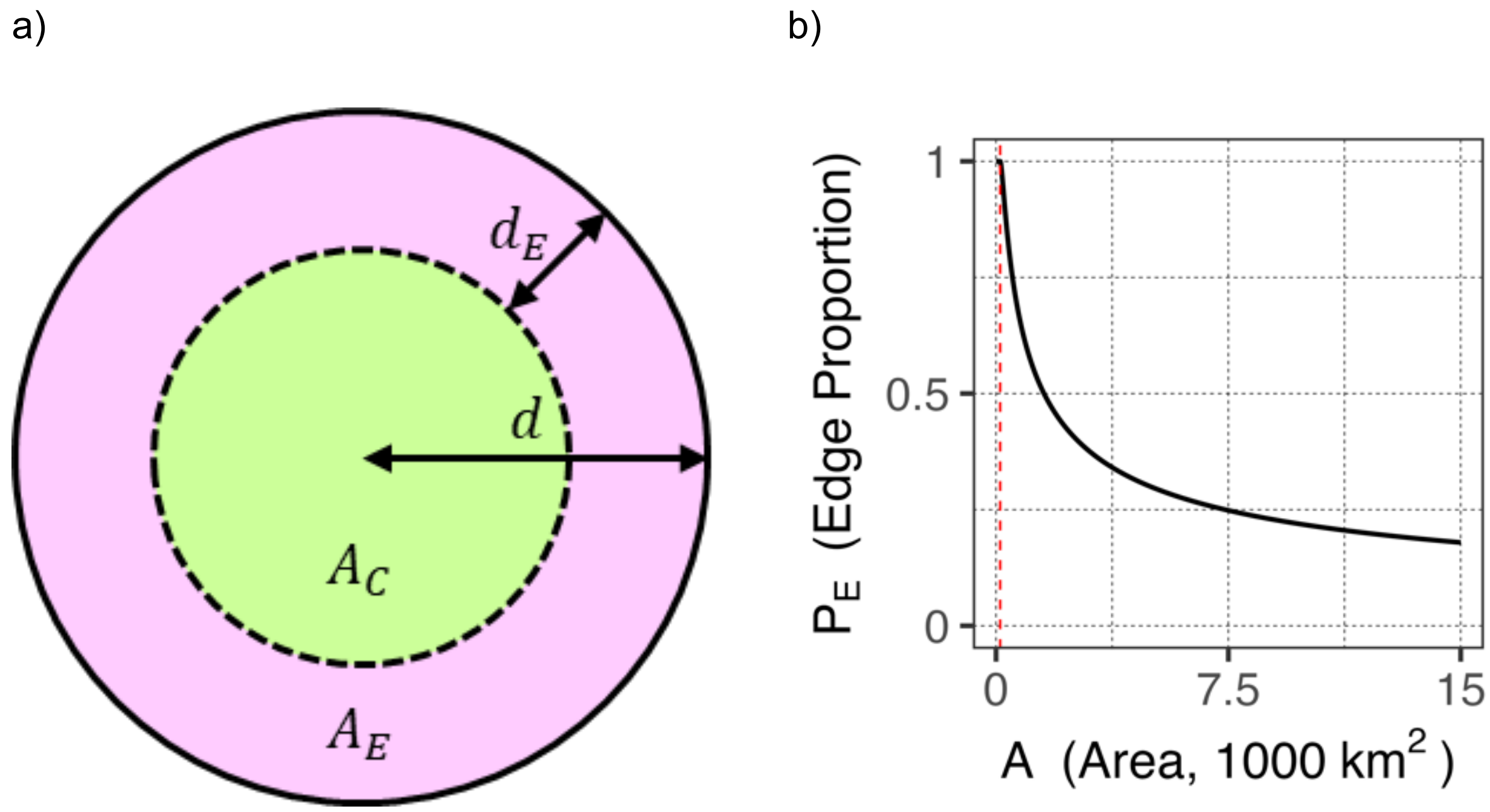

**Figure 1| Modelled protected area shape and threat distribution**
We model an idealised scenario in which a population is distributed evenly across a circular reserve of radius ($\mathbf{d}$) and affected by elevated mortality within its perimeter for distance ($\mathbf{d_E}$). This results in a core area ($\mathbf{A_C}$) which is buffered from the threat occurring in some edge area ($\mathbf{A_E}$). The edge proportion, ($\mathbf{P_E}$) is the proportion of the total reserve area, $\mathbf{A}$, in the edge. The red dashed line shows the area at which core habitat is first established in the reserve.

## Methods

To illustrate our theory, we focus on a population threatened by snare-based poaching, an urgent threat impacting the persistence of wild animal populations across the globe (Bennett et al., 2002, Maxwell et al., 2016). Snare hunting (the use of wire traps to capture animals) is most common in accessible parts of habitat, although it may extend many kilometres from access points (Abernethy et al., 2013, De Souza-Mazurek et al., 2000, Laurance et al., 2000, Laurance et al., 2008, Critchlow et al., 2015). Therefore, it is an ideal threatening process to demonstrate the importance of habitat edges on the benefits of expanding, or better managing, protected areas.

As an example, we consider a population of Peter's Duiker (*Cephalophus callipygus*) threatened by snare poaching. The decision-maker can protect the species by allocating $x_L$ resources towards more land, expanding a protected area, and $x_E$ resources towards better managing the protected area (in the "edge"), through snare removal patrols. To link resource allocation to reducing extinction risk, we adapt a model for mean time to extinction, $M,$ as a function of growth rate, carrying capacity, and environmental stochasticity (Equation 1) (Lande, 1993). We assume poaching reduces the growth rate via direct mortality and that habitat area increases carrying capacity (McCarthy and Possingham 2007). We then optimize investments in protected area expansion and snare removal patrols to maximize mean time to population extinction, highlighting both the nature of the trade-off and the role of initial conditions.

*Population extinction model*

In the following sections we outline our model of how interventions affect extinction risk. We start by assuming mean time to population extinction $M$ is approximated by the function of carrying capacity $K$, mean population growth rate $\bar{r}$, and variance in growth rate $\sigma^2$,

**Equation 1**

$$M = \begin{cases} \frac{2}{\sigma^2 b}\left(\frac{K^b - 1}{b} - lnK\right) & for \quad b \neq 0 \\ \frac{(lnK)^2}{\sigma^2} & for \quad b = 0 \end{cases},$$

where $b = \frac{2\bar{r}}{\sigma^2} - 1$. The above formula assumes an initial population size equal to the population carrying capacity, $K$, which then changes through time diffusively via both average intrinsic population growth, at rate $\bar{r}$, and environmental fluctuations in growth rate, with variance $\sigma^2$. Extinction is assumed to occur once the population reaches a size of one individual (see Lande, 1993, for a full derivation). Note this formula incorporates the fact that environmental stochasticity reduces the long run growth rate of a population below $\bar{r}$, because favourable times do not completely compensate for times when conditions are unfavourable. The true long-run growth rate, accounting for stochasticity, is $\bar{r} - \sigma^2/2$, which is zero when is $\bar{r} = \sigma^2/2$, or equivalently, $b = 0$ (Lande 1993). Therefore, when $\bar{r} > 0$ but $b < 0$, there is still a deterministic drive towards extinction solely due to environmental stochasticity. Note the reason why the mean time to extinction formula needs to be defined piecewise is because the first line is undefined when $b = 0$, but its limit as $b$ approaches zero can be shown to be equal to the second line (see Lande, 1993, for a full derivation).

We assume that all habitat is of equal quality, and that carrying capacity in the absence of poaching is therefore the product of the population density at carrying capacity, $d_K$, and habitat area $A$, $K = d_K A$.

We further assume that hunting pressure supresses mean population growth rate $\bar{r}$ (Robinson and Bennett, 2000, Watson et al., 2013), via additional mortality in the habitat edge (Diamond, 1984, 1989). This mortality is the product of the per capita harvest mortality of individuals exposed to hunting, $m$, and the proportion of the population that is hunted $P_E$. Therefore, $\bar{r} = ln(e^{\bar{r}_C} - mP_E)$, where $\bar{r}_C$ is the baseline population growth rate under no poaching (i.e. the growth rate in the "core").

To determine the per capita mortality, $m$, we introduce a parameter $q$, the catchability of individuals by snares. As $q$ increases, snares kill more individuals. We assume the probability that an individual is not caught, given that a snare is present, decays exponentially with the product of the catchability of a single snare, $q$, and snare density, $S$, in hunted edge habitat. The complement of this value yields the probability that an individual is caught by at least one snare, $m = 1 - e^{-qS}$. To determine $P_E$, we assume that individuals are passively uniformly distributed across the habitat patch and not lower in the edge due to offtake and active snare avoidance. We also assume that the habitat patch is circular in shape, which means $P_E$ is equal to the proportion of habitat that is the hunted edge. for a circular reserve, reserve area is determined by the reserve radius, $d$. The proportion of the reserve that is subject to hunting depends on hunting depth $d_E$, and the distance that snare hunters are active within the edge of the reserve. When $d_E \geq d$, hunters are active across the entire reserve and $P_E = 1$. However, when $d_E < d$, a core of unhunted habitat is established and $P_E$ is equal to the proportion of the reserve that is not core. The radius of the habitat core is equal to $d - d_E$ and therefore the habitat core area, $A_C$, is calculated as $A_C = \pi(d - d_E)^2$. Subtracting this from reserve area, given by $A = \pi d^2$, yields habitat edge area, $A_E$, and dividing this by $A$ yields $P_E$.

Substituting our formulation for additional snare mortality per capita and proportion hunted edge, $P_E$, into intrinsic rate of increase for a population subject to snare hunting in the reserve edge yields,

**Equation 2**

$$\bar{r} = \begin{cases} ln\left(e^{\bar{r}_C} - (1 - e^{-qS})\right) & for \quad A \leq \pi {d_E}^2 \\ ln\left(e^{\bar{r}_C} - (1 - e^{-qS})\left(\frac{2d_E}{\sqrt{\frac{A}{\pi}}} - \frac{\pi {d_E}^2}{A}\right)\right) & for \quad A > \pi {d_E}^2 \end{cases}.$$

To summarize, equation 2 says when the reserve is small (first line), there is no core; the whole reserve is edge, exposed to poaching, and the growth rate is the baseline growth rate, $\bar{r}_C$, reduced by offtake from poaching. When the reserve is large enough (second line), the offtake is multiplied by the ratio of area in the edge, $P_E$ (the bracketed term). It is easy to show that when there is no poaching, $S = 0$, equation 2 reduces to $\bar{r} = \bar{r}_C$, as expected. The exponential and the natural log come from the fact that growth rates are typically reported in continuous time in the literature and poaching mortality is typically reported as proportions or numbers of individuals harvested over a discrete time interval. The exponential and natural log just make these terms comparable such that they can be combined (see Hunting Susceptibility in the appendix for full derivation).

Under this model, increasing snare density, $\boldsymbol{S}$, drives exponential decline in population growth rate. Where reserve radius exceeds hunting depth, increasing area ($\boldsymbol{A}$) drives an increase in the mean population growth rate towards a plateau. When hunting depth is increased, the response of population growth rate to increasing area is supressed.

*Conservation Actions*

Investment in reserve expansion, $x_L$, increases carrying capacity through increased habitat, and also increases $\bar{r}$, through its effect of on hunting depth. Investment in snare patrols (management), $x_E$, only increases $\bar{r}$ through reduced mortality (Equation 2). Here, we define these relationships mathematically allowing us to input them into our population extinction model.

*Expanding the Protected Area*

Protected area expansion impacts mean time to population extinction via two causal pathways (Figure 2).

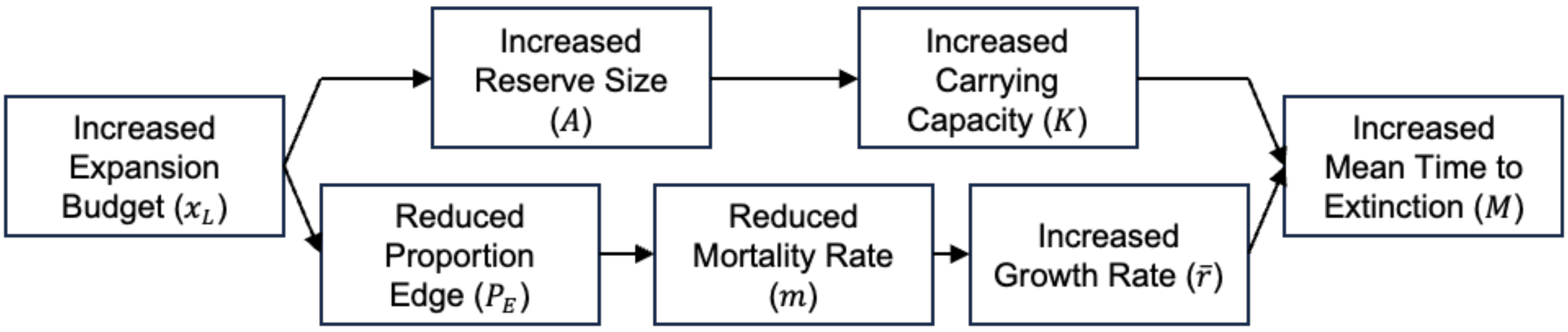

**Figure 2| Causal chain from investment in snare removal and extinction risk**
Investment in snare removal patrols impacts mean time to population extinction via reduced snare density, reduced per capita mortality rate, increased intrinsic rate of population increase.

We assume a fixed annual rent (cost) of new area per unit area, $c_L$ such that $K$ is a linear function of the investment in renting more land, $x_L$. Under zero investment in protected area expansion ($x_L = 0$), $K = d_K A_0$. A further investment in expansion, $x_L$, leads to

**Equation 3**

$$K = d_K \left( A_0 + \frac{x_L}{c_L} \right).$$

Additionally, $x_L$ drives $\bar{r}$ upwards via its negative impact on $P_E$. When $x_L = 0$, $\bar{r}$ is defined by equation 2. But when, $x_L > 0$ , $\bar{r}$ is given by

**Equation 4**

$$\bar{r} = \begin{cases} ln\left(e^{\bar{r}c} - (1 - e^{-qS})\right) & for \quad A_0 + \dfrac{x_L}{c_L} \leq \pi {d_E}^2 \\ ln\left( e^{\bar{r}c} - (1 - e^{-qS}) \left( \dfrac{2d_E}{\sqrt{\dfrac{A_0}{\pi} + \dfrac{x_L}{\pi c_L}}} - \dfrac{\pi {d_E}^2}{A_0 + \dfrac{x_L}{c_L}} \right) \right) & for \quad A_0 + \dfrac{x_L}{c_L} > \pi {d_E}^2 \end{cases}.$$

*Managing snare density*

Investment in snare removal patrols, $x_E$, increases mean time to population extinction via its negative effect on snare density $S$ (in equation 4).

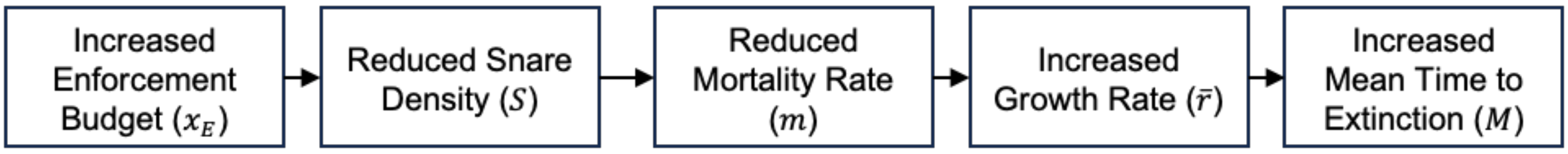

**Figure 3| Causal chain from investment in snare removal to extinction risk**
Investment in snare removal patrols impacts mean time to population extinction via reduced snare density, reduced per capita mortality rate, increased intrinsic rate of population increase.

This reduction in $S$ mitigates the risk of deterministic extinction via its impacts on mortality rate $m$ and, subsequently, population growth rate $\bar{r}$. To model the relationship between $x_E$ and $S$, we assume that rangers move randomly and remove snares, at no cost, when found. The probability of not finding any given snare given that it is present $Pr_{NF|P}$ can be expressed as an exponential decay function with an exponent equal to the product of $x_E$ and patrol efficacy, $s$ (Hauser and McCarthy, 2009). Therefore, the expected snare density $S$ in the plot after a ranger patrol visit is the product of the initial number of snares and $Pr_{NF|P}$.

The annual investment in snare removal patrols is distributed across multiple visits and multiple sites. We assume a regime in which $x_E$ is allocated evenly across the hunted habitat edge at a rate of one visit every $c$ days, yielding an allocation of $x_{EV}$ for a single visit to a single site, which can be expressed as $x_E$ divided by area of hunted edge habitat $P_E A$ and number of visits per year $(365/c)$,

$$x_{EV} = x_E \frac{c}{365\pi d_E \left(2\sqrt{\frac{A}{\pi}} - d_E\right)}.$$

This results in a negative impact of increased area $A$, whereby it drives rapid decline in $x_{EV}$ and an increase in the number of snares left in the habitat. The combined impact of $x_E$ and $x_L$ on $\bar{r}$ is given by substituting, $S = S_0 e^{-s x_{EV}}$ into equation 4.

*Baseline Parameter Values*

We applied our framework to populations of Peter's Duiker (*Cephalophus callipygus*) hunted in edge habitats of protected areas in the Congo Basin of Africa. We obtained estimates of mean population growth rate, $\bar{r}$, variance in mean population growth rate, $\boldsymbol{\sigma^2}$, and population carrying capacity, $K$, from previously reported field data (Barychka et al. 2020). These were summarised from field estimates of the variables themselves and allometric relationships with field estimates of body mass (Barychka et al. 2020).

We calculated hunting susceptibility, $q$, to be 0.2856 $km^2$ $snare^{-1}$ $year^{-1}$ based on published field data on the hunting practices of the Baka peoples in Nki National Park, Cameroon (Yasuoka, 2006, see appendix for detailed derivation). We used a baseline snare density value of 22.5 snares per $km^2$, as this is the mean of four studies in the Congo Basin, with a range of 3.6 – 56. Across four studies in Zambia, Zimbabwe and Kenya the maximum distance travelled by hunters from protected area edges varied from 2-10 km, with an average of 6.5 km (Table 5.3).

We produced plausible values for patrol efficacy, $s$, by applying experimentally derived measures of snare detectability for 2 km of search effort in 1 $km^2$ plots (O'Kelly et al., 2018) and cost per unit patrol effort at Virunga and Kahuzi-Biéga National Park's for the period 2005-2008 (Mubalama-Kakira, 2010) to an exponential decay function (Hauser and McCarthy, 2009). See appendix for detailed derivation.

| | Variable | Baseline Value | Data |
|---|---|---|---|
| Biological Traits | Mean rate population growth, $\bar{r}$ | 0.44 | Barychka et al. (2020) |
| | Variance in rate of population growth, $\sigma^2$ | 0.0196 | Barychka et al. (2020) |
| | Population density at carrying capacity, $d_K$ | 9.7 $km^{-2}$ | Barychka et al. (2020) |
| | Hunting susceptibility, $q$ | 0.2856 $km^2$ $snare^{-1}$ $year^{-1}$ | Yasuoka 2006 |
| Threatening Processes | Hunting Depth, $d_E$ | 6.5 km | Mean value from: Wato et al. (2006), Lindsey et al. (2012), Watson et al. (2013), Kimanzi et al (2015) |
| | Initial snare density, $S_0$ | 22.5 km$^{-2}$ | Mean Value from: Colell et al (1994), Noss (1995), Fa and Garcia Yuste (2001), and Yasuoka (2006) |
| Conservation cost-efficacy | Agricultural Opportunity cost (expansion cost), $c_L$ | 549 USD km$^{-2}$ | Naidoo and Iwamura, 2007 |
| | Snare search efficacy, $s$ | 0.0377 USD$^{-1}$ | Calculated from mean Value from O'Kelly et al. (2018) and Mubalama-Kakira (2010) |

**Table 1| Summary of parameter values**

## Results

The optimal choice between reserve expansion and increasing snare removal patrols depends on initial snare density, $S_0$, and area, $A_0$ (Figures 4-6). for the baseline parameterisation, the species quickly becomes extinct (< 25 years) in protected areas that are both smaller than 100 km$^2$ and contain more than 10 snares per km$^2$ (Figure 4a). As area increases over 132.7 km$^2$ (i.e. $\pi {d_E}^2$), core habitat forms, and extinction risk reduces (increased mean time to extinction, Figure 4a). Protected areas larger than 1,000 km$^2$ create resilient populations regardless of the amount of poaching on the edge, due to a large core of undisturbed habitat (pink region in the upper portion of Figure 4a). Reductions in snare density can also reduce extinction risk. for reserves with less than two snares per km$^2$ in the edge, the population is expected to persist for any reasonable reserve size (pink region in the left portion of Figure 4a). for small reserves, the species go from near-certain persistence to rapid extinction for a small window of snare densities between 2 and 3 snares per km$^2$, after which extinction is expected to occur within 50 years or less (Figure 4a).

When poaching only occurs in the edge, protected area expansion is usually a better investment than snare removal (Figure 4b). for large, protected areas, it is always optimal to expand the protected area (Figure 4b). This is because large, protected areas have a core of unpoached habitat, and investing in expansion expands the core faster than the edge (Fig 1b). This can be seen in Figure 4b, where the black horizontal line represents the reserve size required to create core habitat for a hunting depth of 6.5 km. In small reserves, snare removal is optimal for the same small window of intermediate snare densities where extinction risk transitions between near-certain persistence to rapid extinction.

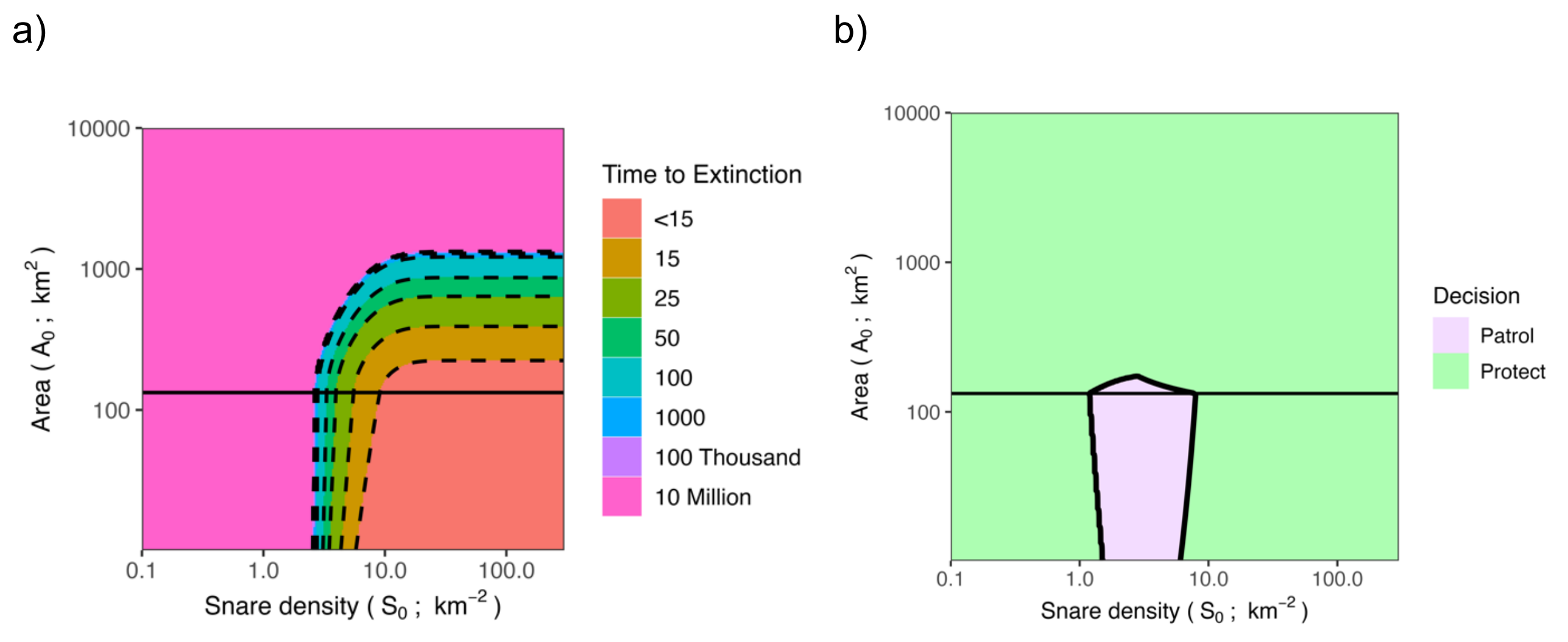

**Figure 4| Optimal action according to initial snare density ($S_0$) and initial area of the protected area ($A_0$)** **a)** Mean time to extinction (colour) as a function of snare density and initial area. Large reserves or reserves with low snare density guarantee persistence (M > 10 million years, pink region on the left). Small reserves with high snare density lead to near immediate extinction (M < 15 years, orange region in the bottom right). **b)** Optimal strategy as a function of initial snare density and area. The horizontal black line in both a) and b) corresponds to the area required to sustain a threat free core given a poaching depth of 6.5 km. Protected area expansion is favoured among protected areas that are large enough to sustain a snare free core (green region above the horizontal black line in b). for smaller reserves, snare removal patrols are favoured when extinction risk is between 15 and 10 million years (compare pink region in b with the vertical slice of gold, green and blue colours below the horizontal black line in a).

The above result is robust to changes in several model parameters. When increasing/decreasing population density at carrying capacity, poaching efficacy (i.e. catchability), mean population growth rate, and variance in population growth rate, across several orders of magnitude from the baseline, the shape and size of the green and pink regions in Figure 3b remain similar (See figures S1 – S4 in the supplementary appendix). However, three parameters can substantially affect the optimality of expanding versus better enforcing protected areas.

The search efficacy of snare-removal patrols, $s$, has a large effect on the expansion snare removal trade-off. Increasing the efficacy of patrols means that more snares are removed by the same investment in patrols. This causes patrols to be more efficient across a broader range of initial area and snare density (Figure 5).

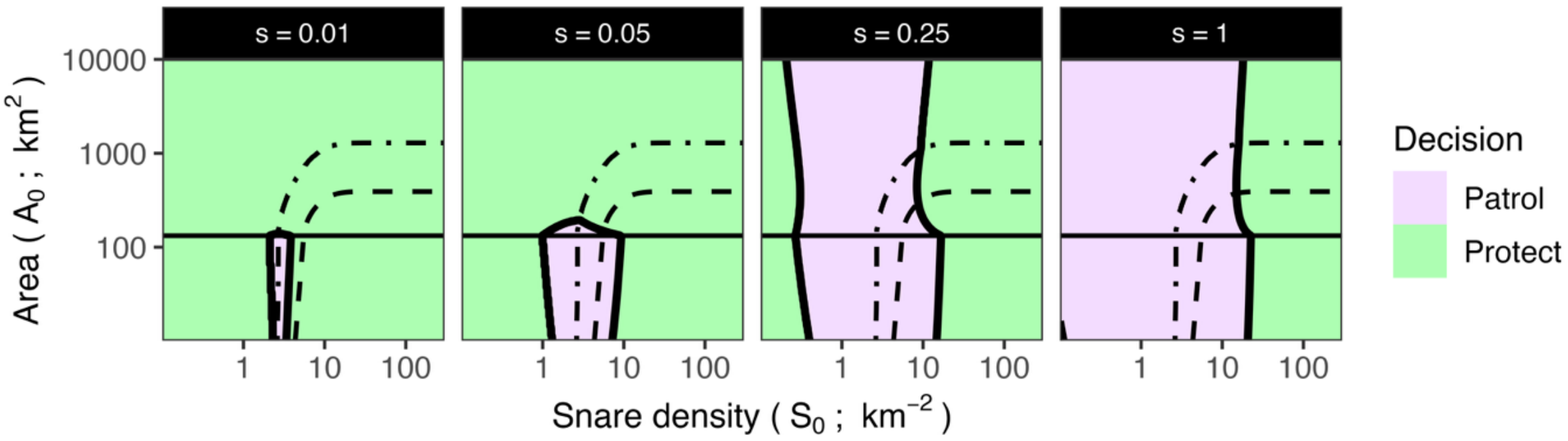

**Figure 5| Effect of snare removal efficacy (s) on optimal decision**
Increasing snare removal efficacy (s) expands the initial conditions under which investment of 1USD in snare removal over expansion is optimal. The dashed and dot-dashed lines represent the protected area size and snare density combinations corresponding to a mean time to extinction of 25 years and 100,000 years respectively. See Figure 4a for how extinction risk changes with respect to area and snare density in the baseline parameterization.

Conversely, greater annual agricultural opportunity cost $c_L$ (a measure of the cost of acquiring land) means that less additional area is secured for a given investment in protected area expansion. This causes protection to become less cost-effective and patrols are again favoured over a broader range of initial areas and snare densities (Figure 5).

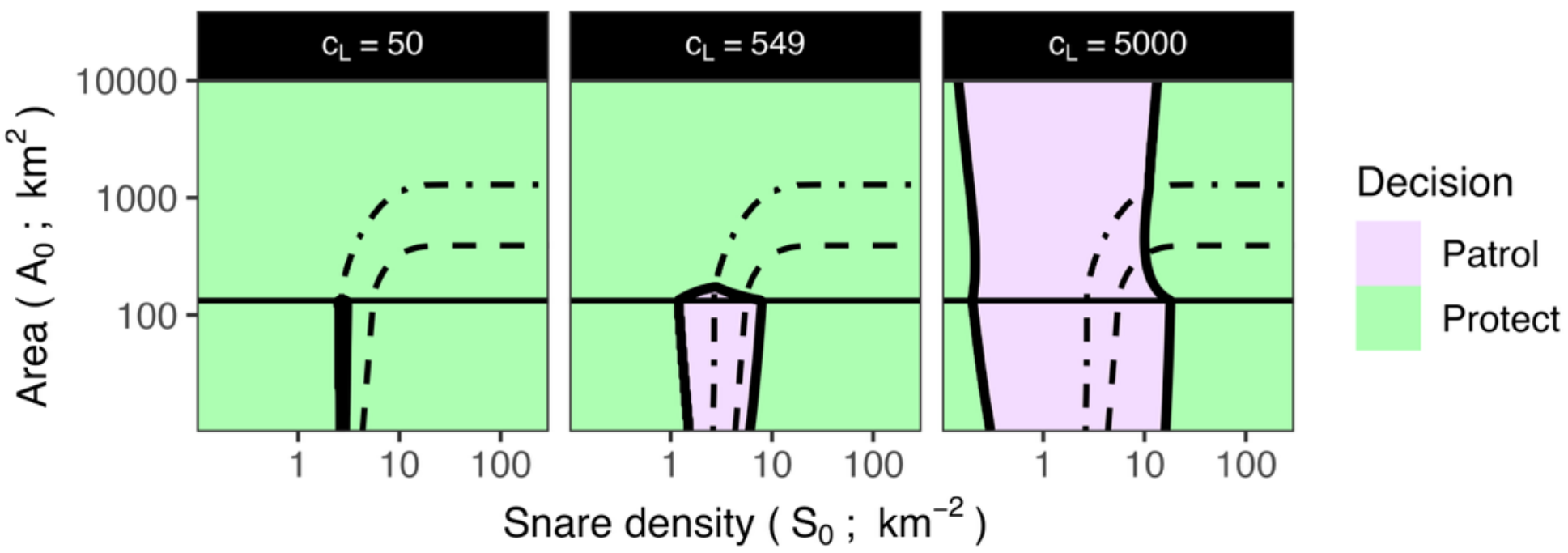

**Figure 6| Effect of annual agricultural opportunity cost ($c_L$) on optimal decision**
Increasing agricultural opportunity cost ($c_L$) expands the initial conditions under which investment of 1USD in snare removal over expansion is optimal. The dashed and dot-dashed lines represent the protected area size and snare density combinations corresponding to a mean time to extinction of 25 years and 100,000 years respectively. See Figure 4a for how extinction risk changes with respect to area and snare density in the baseline parameterization.

Changes in hunting depth $d_E$ also affect the optimal decision (Figure 6). When poachers can travel further into the reserve, the area required to maintain a core of poacher-free habitat increases. This has the effect of expanding pink region to greater values of initial area (compare Figure 6 to Figure 4b) meaning for intermediate snare densities, increasing snare removal patrols can be optimal over reserve expansion even for reserves up to 1,000 $km^2$.

These reserves were initially large enough to sustain a small core for small hunting depths. However, for larger poaching depths the core is lost unless the reserve is substantially larger.

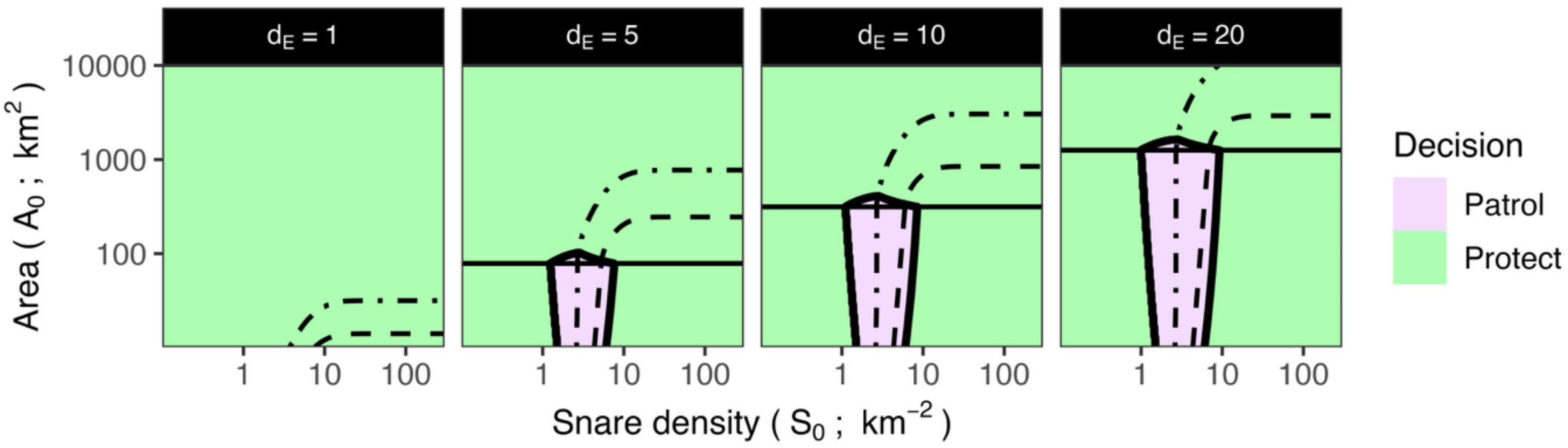

**Figure 7| Effect of hunting depth ($d_E$) on optimal decision**
Increasing hunting depth ($d_E$) stretches the pink region where an additional investment of 1USD in snare removal ($x_E$) over expansion ($x_L$) is optimal. The dashed and dot-dashed lines represent the protected area size and snare density combinations corresponding to a mean time to extinction of 25 years and 100,000 years respectively. . See Figure 4a for how extinction risk changes with respect to area and snare density in the baseline parameterization.

***Case study: Implementing the framework for real, non-circular reserves***

Two of the most important variables, hunting depth and agricultural land opportunity cost are likely to be highly variable in space, depending on the geometry and characteristics of the reserves being considered. To implement our decision framework for protecting *C. callipygus* across a set of real serves we obtained initial area estimates of protected areas from the World Database of Protected Areas (WDPA) (UNEP-WCMC and IUCN, 2023) for all national level, IUCN category I-IV designated reserves containing *C. callipygus*.

The protected areas were irregular in shape, and our model, which assumes that the protected area is circular, is likely to introduce a bias towards protected area expansion because a smaller overall area is hunted, and protected area expansion increases the area of the unhunted core. To reduce this bias, we calculate an adjusted hunting depth, $d_{E\ Adj}$, that represents the distance that hunters would have to be active within the protected area perimeter to cover the same edge area if the reserve was circular (Equation 5). Having previously obtained the extent of protected areas, $A$, we proceeded to obtain an unhunted core area, $A_C$, by creating an internal buffer of 6.5 km, the mean distance of the furthest snare from the edge of four protected areas (Table 5.2). Next, we obtained the radius of a circle of the size of the total area, $d$, and that of a circle of the area of unhunted core, $d_C$. Where $d > d_C$, $d_{E\ Adj}$, is the difference between these values (Equation 5). Where no core remains after the internal buffer is applied, we assigned a high value to $d_{E\ Adj}$ such that protected area expansion never creates unhunted core. This is appropriate given the small investment being considered and the low chance that this would cause a reserve to create new core where none existed previously,

**Equation 5**

$$d_{E\ Adj} = d - d_C = \sqrt{\frac{A}{\pi}} - \sqrt{\frac{A_C}{\pi}}.$$

We produced plausible values for land cost by calculating the mean annual agricultural opportunity cost (Naidoo and Iwamura, 2007) within a 1 km buffer around protected area polygons in the World Database on Protected Areas (UNEP-WCMC and IUCN, 2023) using ArcGIS software and the “maptools” and “raster” packages in R (R Core Team, 2016). The land costs and adjusted hunting depth for each reserve is displayed in figures S6 and S7 in the online supplementary information. Protected areas that are too small to support one individual at carrying capacity are immediately extinct under our model (Lande 1993) and were therefore excluded from the analysis.

## ***Case Study Results***

Protected area expansion is the optimal action for reducing the extinction risk of Peter’s Duiker (*Cephalophus callipygus*) populations in all protected areas within its geographic range (Figure 8). The main takeaway is that search efficacy would need to increase by approximately 116-fold before increasing snare removal patrols would be more beneficial for protecting *C. callipygus* than expansion in half the protected areas (range of 18 - 433-fold multipliers across all 24 protected areas). The protected areas that more strongly favoured expansion and therefore require higher improvements in search efficacy for snare removal to be optimal, tend to be larger and, to a lesser degree, rounder in shape (Figure 8). Additionally, reserves in areas with lower agricultural opportunity cost also tend to have higher multipliers due to the relative cost efficiency of cheap expansion (compare Figure 8 to Figure S7 in the online supplementary appendix). Reserves with the lowest required multipliers were small, without a core (Figure S8) and had a short mean time to extinction (Figure S9).

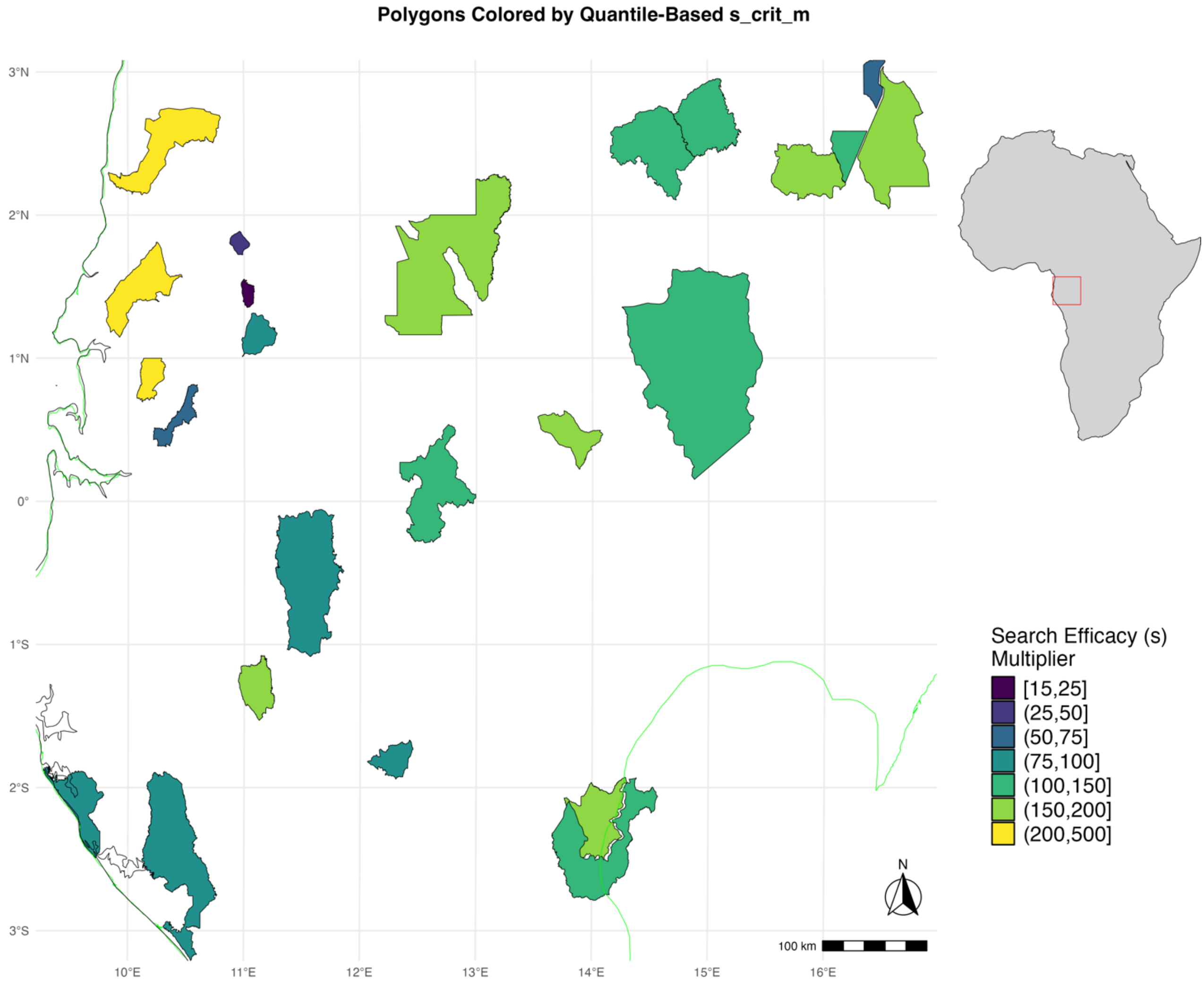

**Figure 8| Search efficacy (s) at which snare removal becomes optimal**
Multiplicative improvement in snare search efficacy required for the optimal investment to switch from expansion to increased snare removal. for example, a multiplier of two would require search efficacy to double for increased snare removal to be the optimal conservation investment. The four yellow reserves with extremely high multipliers in the upper left are due to low agricultural opportunity costs in those areas leading to cheaper expansion (Fig S7). The lowest multipliers are in the small reserves because they do not contain a threat-free core (Fig S8); however, expansion is still favored in these reserves due to low mean times to extinction (Fig S9).

## Discussion

The allocation of scarce conservation resources between expanding protected areas and improving protected area management is a major conundrum in conservation. While several international agreements have called for rapid expansion, previous prioritization studies suggest that expansion without effective management is suboptimal (Possingham et al 2015, Kuempel et al 2016, Adams et al., 2019, Timms and Holden 2023). We showed that the optimal balance between expansion and management is context dependent. In our model, when threats are concentrated near edges, protected areas are large enough to form a threat free core, and management is costly or inefficient, then expanding protected areas is likely a good investment. This result arises because the central core of the protected area expands faster than the edge as the reserve grows.

However, for small, protected areas that do not initially have a core (or potentially if threats can spread into the entire reserve) management will likely often be more cost effective than expansion. While expansion might be numerically optimal for small reserves in some extreme cases where extinction is imminent (mean time to extinction < 15 years) or highly unlikely (mean time to extinction > 1 million years), these scenarios either likely preclude meaningful intervention, or do not require it. Thus, for small reserves where timely action can still make a difference, investing in management is likely more essential than expansion.

In our case study, minimizing the extinction risk of Peter's Duiker in the dense forest of Africa's Congo Basin, expansion consistently outperformed increased snare removal patrols across 24 protected areas. This result was likely due to the fact that all 24 protected areas were either large enough to maintain a core, or (among the smaller reserves) had short mean times to extinction. In these reserves, snare removal would have to become 18- to 433-fold times more cost effective to outperform expansion (median of 116). Therefore, only substantial improvements in patrol strategy or technology could make snare removal competitive. Such improvements are possible. Our analysis assumed snare removal patrols were uniformly distributed across the hunted area. Targeting known tracks or incorporating emerging technologies such as drones (Bondi et al., 2018) or models that predict poaching to optimize patrol deployment (Critchlow et al., 2015, 2017, Gholami et al., 2017, Xu et al., 2020) could substantially improve efficiency. Our framework provides a concrete benchmark: a 116-fold efficacy gain for snare removal to outperform expansion.

Our baseline model assumes patrols only reduce poaching by directly removing snares. If poachers set fewer snares in response to patrols, snare removal becomes more effective. This could be incorporated into our model via a multiplier improving snare removal efficacy. In the case study, poachers would need to set 18 fewer snares, per snare removed by a patrol (snare removal efficacy improved by 18-fold), for patrols to be more cost effective than expansion in at least one reserve. However, in the sensitivity analysis, in Figure 6, the region of parameter space where snare removal is optimal can be substantial, if patrols reduce poaching by five-fold over the baseline value.

Our results are robust to a wide range of parameter values, but they rely on simplifying assumptions. for example, we considered uniform hunting within a fixed hunting depth whereas real-world snare density often declines or peaks at intermediate distances from reserve edges (Kimanzi et al. 2015). These patterns may reflect terrain heterogeneity or poacher responses to patrols. Technological innovations for poachers could also affect future spatial poaching patterns. Another source of spatial heterogeneity is the presence of roads, which may facilitate access and increase threat penetration (De Souza-Mazurek et al., 2000, Laurance et al., 2000, Laurance et al., 2008, Abernethy et al., 2013, Critchlow et al., 2015). We repeated the analysis including roads bisecting protected areas, and this did not qualitatively affect our results although it did improve the relative benefit of snare removal (see appendix). Our model can accommodate complex static poaching distributions by dividing the edge into multiple zones with different snaring intensities to match the details of different examples. Similarly, it could statically account for animals avoiding snares or stress effects (i.e. a “landscape of fear”) via modifications to the population growth rate.

Incorporating dynamic interactions between poachers, wildlife, and rangers poses a greater modelling challenge. for example, in response to increased management poachers may penetrate deeper into reserves (Bode et al 2015) to avoid their snares being removed. This could be incorporated statically by letting hunting depth be a function of snare density, but this modification still misses any further repeated back and forth responses between patrols and poachers. Additionally, poachers may also increase effort when animal populations are more abundant. While such feedbacks are captured in dynamic population-poaching models (e.g. Kuempel et al 2018, Holden and Lockyer 2021, Timms and Holden 2023), they are harder to represent in extinction-risk models like ours. As a result, we assume a static, uniformly distributed population.

The most comprehensive extension would involve spatially explicit, game-theoretic models capturing strategic interactions: poachers seeking prey, animals avoiding poachers, poachers avoiding patrols, and patrols seeking poachers. To date, these dynamics have not been fully incorporated into expand-versus-manage models. Calculating mean time to extinction in such complex models is mathematically and numerically intractable, since it is sensitive to rare instances of long extinction times. One possible way to overcome such challenges is to simulate behavioural models until they reach spatial equilibrium (provided a stable equilibrium exists) and applying the resulting spatial patterns in our static extinction risk model. Alternatively, one could simulate short term extinction probabilities (e.g. over 20 years), which may be tractable and policy relevant. Developing such challenging, strategic, dynamic models remains one of the most important open questions in this field of research.

Our focus on poaching for food using passive snares in dense forest, likely makes our assumptions more realistic than in cases involving highly mobile poachers seeking high value species across open terrain (e.g. oceans or savannas). In such systems, poachers may access entire reserves and respond more dynamically to enforcement. Active interventions like catching and punishing profit-seeking poachers (Kuempel et al. 2018, Holden et al 2019, Holden and Lockyer 2021, Lopes 2024, Kibira et al. 2024) are more likely to violate our assumption of static threats.

Lastly, we assumed the management of a single species. Extending our framework to multiple species could involve minimizing the geometric mean of mean extinction times across a set of species which prioritizes the most vulnerable species and therefore aligns well with biodiversity conservation objectives (Erm et al. 2023, Erm et al. 2024). Such an extension is especially important given concerns around bycatch (Erm et al 2024, Takashina 2024) and opportunistic exploitation (Branch et al 2013, Thurner et al 2021).

A key advance of our framework is its focus on outcomes (minimized extinction risk) rather than intermediate action-based target. Conservation interventions are often selected based on threat reduction or area coverage rather than their impact on species persistence. (Pressey et al. 2007; Tulloch et al. 2015). Our approach can help redirect funds towards the most cost-effective actions even in the absence of empirical evidence on intervention efficacy.  In the absence of such evidence, modelling provides the best available guidance for choosing actions that reduce extinction risk and advance conservation.